\documentclass[review,12pt]{elsarticle}

\usepackage{amsmath,amssymb,mathtools}
\usepackage{booktabs}
\usepackage{array}
\usepackage{graphicx}
\usepackage{enumitem}
\usepackage[hidelinks]{hyperref}

\journal{Knowledge-Based Systems}

\graphicspath{{}{figures/}{prod_case/}}

\newcommand{\maybeincludegraphics}[2][]{%
  \IfFileExists{#2}{\includegraphics[#1]{#2}}{%
    \fbox{\parbox{0.86\linewidth}{Figure file {\ttfamily\detokenize{#2}} not found in this build. The manuscript source will include the figure when the file is supplied.}}}}

\begin{document}

\begin{frontmatter}

\title{Cost-sensitive retraining via posterior learning debt}

\author[airbnb]{Harrison E. Katz\corref{cor1}}
\ead{harrison.katz@airbnb.com}
\cortext[cor1]{Corresponding author.}
\affiliation[airbnb]{organization={Finance Data Science \& Strategy, Airbnb Inc.},
  city={San Francisco}, state={CA}, country={USA}}

\begin{abstract}
Deployed prediction systems are often retrained on fixed calendars, even when
model staleness and retraining burden vary over time. This short communication
formulates retraining for Bayesian prediction systems as a cost-sensitive
predictive-regret decision. The central monitoring state is posterior learning
debt, defined as the Kullback--Leibler divergence from a reference shadow
posterior to the deployed frozen posterior. In the decision layer, a retraining
cost is compared with the expected one-period predictive regret of waiting. A
continuous-severity version retrains when calibrated expected regret exceeds the
retraining cost, while the familiar two-state excess-loss rule is a special
case. The empirical study is an exact-state proof-of-concept in a
synthetic conjugate simulation with warm-started deployed and shadow
normal-inverse-gamma posteriors, separate update, monitoring, and evaluation
batches, lagged deployment actions, expanded baseline grids, and score-unit
sensitivity. Under the primary 75th-percentile
score-unit scaling, an age-adjusted debt-threshold policy improves on tuned
calendar retraining in all 72 non-stable scenario cells and on tuned CUSUM in
58 of 72 cells, with mean relative objectives 0.677 and 0.975, respectively.
Debt-utility and hybrid-utility policies also improve strongly over tuned
calendar retraining, but they do not dominate tuned CUSUM. Median and mean
score-unit sensitivities show the same main calendar result, while the CUSUM
comparison remains policy-dependent. The contribution is a transparent decision
layer for deployed Bayesian prediction systems, not a universal replacement for
drift detection.
\end{abstract}


\end{frontmatter}

\setlength{\bibsep}{0pt}

\section{Introduction}
\label{sec:introduction}

Calendar retraining is easy to schedule, but it is rarely a decision rule. It
can retrain too early, creating compute, validation, governance, and
user-disruption costs, or too late, allowing stale predictions to accumulate.
The right action depends on the predictive value left on the table by waiting
and on the cost of refreshing the deployed model.

This problem sits between several literatures that are usually kept
separate. Forecasting work studies structural breaks, adaptive estimation, and
online model averaging \cite{clements2006breaks,castle2016overview,
pesaran2007selection,giraitis2013adaptive,raftery2010online}. Data-stream work
studies concept drift, adaptive windows, and online changepoint detection
\cite{gama2014survey,bifet2007adaptive,bach2010bayesian,adams2007bocd,
fearnhead2007online}. Cost-sensitive adaptation and deployment work study when
updates are worth their operational burden
\cite{zliobaite2015cost,spiliotis2024update,verachtert2023budget,
mahadevan2024costaware}. Production-ML and operations research emphasize that
technical debt, forecast accuracy, and downstream decision quality need not
move together \cite{sculley2015hidden,breck2017mltest,goltsos2022mindgap,
kourentzes2020optimising,gammelli2022predictive}.

Mahadevan and Mathioudakis \cite{mahadevan2024costaware}, in particular, study
cost-aware retraining for machine learning data streams as an explicit
cost-loss update problem. The present note is complementary. It does not
propose a general retraining optimizer; it asks how a deployed Bayesian system
can use the separation between deployed and shadow posteriors as an operational
state variable, then calibrate the resulting decision layer to predictive
regret. The claim is intentionally bounded: learning debt is a state variable
and decision layer, not a solved dynamic control problem, not a general
stream-drift benchmark, and not a universal replacement for drift detection.

\section{Posterior learning debt}
\label{sec:debt}

Let $Y_{t+h}$ be the outcome predicted at horizon $h$ from origin $t$, and let
$\mathcal F_t$ be the available information. The unknown target is the true
conditional distribution $F_t^0(\cdot\mid\mathcal F_t)$, which need not belong
to the model. A Bayesian prediction model specifies sampling laws
$\{p_\theta(\cdot\mid\mathcal F_t):\theta\in\Theta\}$. After data through time
$t$, the learner has a posterior measure $\Pi_t(d\theta)$ over model quantities.
The posterior is not $F_t^0$; it induces the posterior predictive distribution
\begin{equation}
Q_t(A\mid\mathcal F_t)=\int_\Theta p_\theta(A\mid\mathcal F_t)\,\Pi_t(d\theta).
\label{eq:predictive}
\end{equation}

Let $\tau(t)$ be the most recent production retraining time. The deployed
posterior is $\Pi_t^{\mathrm{dep}}=\Pi_{\tau(t)}$. Let
$\Pi_t^{\mathrm{ref}}$ be the reference posterior under the monitoring update,
such as continuous Bayesian updating or a shadow learner. Posterior learning
debt is
\begin{equation}
\mathcal D_t^{\mathrm{post}}
=D_{\mathrm{KL}}\left(\Pi_t^{\mathrm{ref}}\,\middle\|\,\Pi_t^{\mathrm{dep}}\right).
\label{eq:postdebt}
\end{equation}
The deployed posterior is the approximation, and the reference posterior is the
target implied by the chosen update rule. The corresponding predictive debt is
\begin{equation}
\mathcal D_t^{\mathrm{pred}}(\mathcal F_t)
=D_{\mathrm{KL}}\left(Q_t^{\mathrm{ref}}(\cdot\mid\mathcal F_t)\,\middle\|\,
Q_t^{\mathrm{dep}}(\cdot\mid\mathcal F_t)\right).
\label{eq:preddebt}
\end{equation}
Thus posterior debt measures separation in the learning state, while predictive
debt measures the induced separation in forecast distributions.

\section{Cost-sensitive retraining rule}
\label{sec:rule}

The most direct operational rule compares the cost of retraining with the
predictive regret of waiting. Let $S(Q,y)$ be a predictive score to be maximized,
such as log score. If $Q_t^{\mathrm{ref}}$ is the shadow predictive distribution
and $Q_t^{\mathrm{dep}}$ is the deployed predictive distribution, define the
one-period positive predictive regret of the deployed model as
\begin{equation}
r_t=\max\{0,\,S(Q_t^{\mathrm{ref}},Y_t)-S(Q_t^{\mathrm{dep}},Y_t)\}.
\label{eq:regret}
\end{equation}
Let $\lambda>0$ be the retraining cost in the same score units. If
$\widehat r_t=\mathbb E(r_t\mid m_{1:t})$ is a calibrated estimate of the
predictive regret of waiting one more period, the one-step decision is
\begin{equation}
\text{retrain if } \widehat r_t>\lambda.
\label{eq:continuous-rule}
\end{equation}
This is the continuous-severity version of the same cost-sensitive logic.

The familiar two-state rule is a special case. Let $Z_t=1$ denote actionable
staleness and let $\rho_t=\Pr(Z_t=1\mid m_{1:t})$. If retraining when $Z_t=0$
incurs excess loss $c_{\mathrm{churn}}$ and waiting when $Z_t=1$ incurs excess
loss $c_{\mathrm{wait}}$, then the standard Bayes action under posterior
expected loss \cite[Chapter~2]{berger1985statisticaldecision} is
\begin{equation}
\rho_t>\frac{c_{\mathrm{churn}}}{c_{\mathrm{churn}}+c_{\mathrm{wait}}},
\qquad\text{or}\qquad
\frac{\rho_t}{1-\rho_t}>\frac{c_{\mathrm{churn}}}{c_{\mathrm{wait}}}.
\label{eq:binary-threshold}
\end{equation}
The simulation below uses the continuous predictive-regret form in
\eqref{eq:continuous-rule}. A full dynamic stopping formulation would add
continuation value and can change the boundary.

\section{Deployment-calibrated simulation}
\label{sec:simulation}

The simulation uses conjugate Bayesian linear regression because it yields exact
updates and exact Kullback--Leibler divergence between deployed and shadow
normal-inverse-gamma (NIG) posteriors. Before monitoring begins, both deployed
and shadow posteriors are warm-started with $300$ stable observations. During
the $T=200$ period monitoring horizon, each period generates separate update,
monitoring, and evaluation batches. Five observations update the shadow
posterior, five observations compute monitoring features, and forty independent
holdout observations evaluate predictive regret. Period-$t$ actions affect the
deployment from period $t+1$ onward.

The data-generating model is
\begin{equation}
Y_t=\beta_tX_t+\varepsilon_t,
\qquad \varepsilon_t\sim\mathcal N(0,\sigma_t^2),
\qquad X_t\sim\mathcal N(0,1),
\end{equation}
under NIG hyperparameters $(\mu_0,\kappa_0,\alpha_0,\beta_0)=(0,1,2,1)$. The
regimes are no shift, abrupt coefficient shift, variance shift, and gradual
drift. After a burn-in period, an abrupt coefficient shift adds a single
$\mathcal N(0,2^2)$ perturbation to $\beta_t$; a variance shift multiplies
$\sigma_t^2$ once by a draw from $\mathrm{Uniform}(3,6)$; and, after
gradual-drift onset, $\beta_t$ receives period-by-period
$\mathcal N(0,0.15^2)$ increments. Shift probabilities are
$\{0.02,0.05,0.10,0.20\}$.

The primary objective for policy $p$ is
\begin{equation}
J_p(\lambda)=\lambda N_{p,\mathrm{retrain}}+
\sum_{t=1}^T \max\{0,\ell_t^{\mathrm{shadow}}-\ell_{p,t}^{\mathrm{dep}}\},
\label{eq:objective}
\end{equation}
where $\ell_t^{\mathrm{shadow}}$ and $\ell_{p,t}^{\mathrm{dep}}$ are evaluation
log scores on the independent holdout batch. The retraining cost is
$\lambda_\kappa=\kappa s_0$, with
$\kappa\in\{0.1,0.25,0.5,1,2,4\}$. The primary score unit is the 75th percentile
of positive one-period predictive regret in calibration data,
$s_0=0.0807$. Sensitivity analyses use the median, $0.0125$, and the mean,
$0.1490$.

A key practical issue is that raw posterior KL can grow under stationarity as
the shadow posterior keeps learning while the deployed posterior is frozen. The
simulation therefore age-adjusts debt. Let $a_t=t-\tau(t)$ be deployment-spell
age and let $d_t=\sqrt{D_{\mathrm{KL}}(\mathrm{NIG}_t^{\mathrm{shadow}}\|\mathrm{NIG}_{\tau(t)}^{\mathrm{dep}})}$.
A stable no-shift calibration sample estimates
\begin{equation}
\mu_0(a)=\mathbb E_0(d_t\mid a_t=a),
\qquad
\tilde d_t=\frac{d_t-\widehat\mu_0(a_t)}{\widehat\sigma_0}.
\end{equation}
The adjusted signal $\tilde d_t$ is the main learning-debt monitor.

The comparison includes three debt policies. The debt-threshold policy retrains
when the age-adjusted debt signal exceeds a threshold tuned on calibration
paths. The debt-utility policy fits a simple calibration model for
$\log(1+r_t)$ using positive age-adjusted debt and spell age, then retrains when
the calibrated expected regret exceeds $\lambda$. The hybrid-utility policy adds
monitoring score gap, posterior-mean gap, residual exceedance, and raw debt to
that calibration. Baselines are tuned calendar retraining, tuned CUSUM on the positive score gap
\cite{page1954cusum}, tuned alarm-only debt thresholds, always-retrain, and
never-retrain. The calendar and CUSUM grids are deliberately wide and are listed
in the replication script. CUSUM is the deliberately strong score-based
comparator: it is sequential, tuned on calibration paths, reset after retraining,
and driven by the positive predictive-score gap that also enters the
predictive-regret objective. Windowing and classification-error detectors such
as ADWIN, Page-Hinkley, DDM/EDDM, and Bayesian online changepoint methods are
natural future comparators \cite{page1954cusum,bifet2007adaptive,gama2014survey,adams2007bocd},
but in this forecasting setting they require additional choices about how to map
probabilistic forecasts, log-score gaps, or posterior states into error streams,
window statistics, hazards, and retraining actions. The simulation is therefore
an exact-state proof-of-concept for the debt signal and decision layer, not a
comprehensive drift-detector benchmark. Calibration uses 20 paths per scenario
cell to fit debt adjustment and utility models, 8 paths per scenario cell to tune
policy parameters, and 100 independent held-out paths per scenario cell for
evaluation.

\section{Results}
\label{sec:results}

Table~\ref{tab:primary-results} reports the primary 75th-percentile score-unit
comparison across the 72 non-stable scenario cells. The strongest debt policy is
the simple age-adjusted debt-threshold rule. It improves on tuned calendar
retraining in all 72 cells, with mean relative objective 0.677, and is
competitive with tuned CUSUM, winning 58 of 72 cells with mean relative
objective 0.975. The regression-style utility policies also improve strongly on
tuned calendar retraining, but they are not better than tuned CUSUM in this
simulation. The intervals in Table~\ref{tab:primary-results} quantify Monte
Carlo uncertainty conditional on the simulated design, not uncertainty about
richer model classes or other data-stream environments.

\begin{table}[t]
\centering
\caption{Primary v13d results under 75th-percentile score-unit scaling. Each row summarizes 72 non-stable cells: three shift families, four shift probabilities, and six cost ratios. Mean relative objectives use paired held-out paths; brackets give paired bootstrap 95 percent Monte Carlo intervals, conditional on the v13d synthetic design, obtained by resampling test paths within each scenario cell. Median and interquartile range are across the 72 cell-level relative objectives. Relative objective below one favors the debt policy.}
\label{tab:primary-results}
\scriptsize
\begin{tabular}{@{}>{\raggedright\arraybackslash}p{0.24\linewidth}>{\raggedright\arraybackslash}p{0.15\linewidth}ccc@{}}
\toprule
Policy & Benchmark & Wins & Mean rel. [95\% CI] & Median rel. [IQR] \\
\midrule
Age-adjusted debt threshold & Tuned calendar & 72/72 & 0.677 [0.674, 0.681] & 0.670 [0.641, 0.705] \\
Age-adjusted debt threshold & Tuned CUSUM    & 58/72 & 0.975 [0.973, 0.977] & 0.964 [0.945, 0.986] \\
Debt utility                 & Tuned calendar & 67/72 & 0.801 [0.795, 0.809] & 0.787 [0.721, 0.881] \\
Debt utility                 & Tuned CUSUM    &  6/72 & 1.158 [1.150, 1.168] & 1.133 [1.037, 1.225] \\
Hybrid utility               & Tuned calendar & 69/72 & 0.795 [0.790, 0.802] & 0.809 [0.701, 0.893] \\
Hybrid utility               & Tuned CUSUM    &  2/72 & 1.148 [1.141, 1.157] & 1.089 [1.049, 1.220] \\
\bottomrule
\end{tabular}
\end{table}

The shift-family decomposition is consistent with that summary. Under the
primary score unit, the age-adjusted debt-threshold policy beats tuned calendar
in 24 of 24 abrupt coefficient-shift cells, 24 of 24 gradual-drift cells, and
24 of 24 variance-shift cells, with mean relative objectives 0.609, 0.733, and
0.689. Against tuned CUSUM, the corresponding wins are 20, 18, and 20 of 24,
with mean relative objectives 0.980, 0.994, and 0.950. These are not large
CUSUM margins, but they show that a debt state can remain competitive with a
strong score-based detector after both policies are tuned.

The score-unit sensitivity is important. With the mean positive-regret score
unit, the debt-threshold policy again beats tuned calendar in all 72 non-stable
cells and beats tuned CUSUM in 63 of 72, with mean relative objectives 0.639 and
0.965. With the median positive-regret score unit, retraining is very cheap: the
calendar and CUSUM baselines tune toward aggressive boundary values, and the
debt-threshold policy beats tuned calendar in 48 of 72 cells and tuned CUSUM in
26 of 72, with mean relative objectives 0.895 and 1.049. Thus the main calendar
result is robust to score-unit scaling, while the CUSUM comparison is sensitive
to the cost scale and policy class.

Stable no-shift behavior is also informative. Under the primary 75th-percentile
score unit, the debt-threshold policy averages 1.39 retrains and objective
0.565 across no-shift cells. Tuned calendar averages 46.83 retrains and
objective 2.773, always-retrain averages 199 retrains and objective 21.064, and
never-retrain averages objective 0.880. Tuned CUSUM remains slightly lower than
the debt-threshold policy in no-shift cells, with objective 0.550 and 8.06
retrains. The age adjustment therefore prevents ordinary stable-spell learning
from becoming automatic retraining, while not eliminating all false-alarm trade-offs.

These results change the interpretation of the empirical contribution. The
paper does not show that a regression-style utility map dominates CUSUM. It
shows that posterior learning debt, after adjustment for deployment age and
calibration on predictive-regret costs, is a useful state variable for replacing
fixed retraining cadence. The simplest calibrated threshold on that debt signal
is the strongest policy in this experiment.

\section{Discussion}
\label{sec:discussion}

Posterior learning debt gives a deployed Bayesian prediction system a compact
state variable for deferred updating. Coupled with a predictive-regret cost
scale, it converts retraining from a calendar convention into a cost-sensitive
decision. The v13d simulation is deliberately more deployment-calibrated than a
binary drift-detection exercise: it warm-starts the model, separates update,
monitoring, and evaluation data, applies actions with a one-period lag, tunes
baselines on calibration paths, evaluates on held-out paths, and checks
score-unit sensitivity. It should nevertheless be read as an exact-state
proof-of-concept in a synthetic conjugate model, not as a general empirical
benchmark for all stream-drift or retraining algorithms.

The results support a bounded but useful claim. Age-adjusted debt-thresholding
strongly improves over tuned fixed-cadence retraining and over trivial
always-retrain and never-retrain extremes. It is also competitive with tuned
CUSUM under the primary and mean score-unit scalings, although the CUSUM
comparison is not uniformly dominant and becomes unfavorable under the very low
median score-unit cost. The utility-calibrated debt and hybrid policies improve
on tuned calendar retraining but do not outperform CUSUM. That pattern is
substantively useful: it says that the learning-debt signal matters, but that
simple, transparent calibration can beat a more complicated regression-style
utility map in this stylized setting.

The limitations are clear. The experiment is conjugate and synthetic by design.
The exact NIG KL is available only because the model is analytically tractable,
and the simulation gives direct access to a shadow posterior that may be harder
to maintain in richer production models. The age adjustment is calibrated from
stable simulation paths, and the tuning objective depends on a selected
score-unit scale. The decision rule is one-step, not a fully solved
optimal-stopping policy. The next step is to replace exact posterior KL with
calibrated predictive-debt or proper-score diagnostics in richer model classes,
then compare those policies against broader online change detectors, including
ADWIN-style windows, Page-Hinkley and DDM/EDDM variants, Bayesian online
changepoint detectors, and dynamic retraining rules
\cite{gama2014survey,bifet2007adaptive,adams2007bocd,fearnhead2007online}.

\section*{Declarations}

\noindent\textbf{Data and code availability.} The reproducible simulation uses
synthetic data generated by the public R script
\texttt{learning\_debt\_sim\_sensitivity.R}. The R directory for the replication script is available at
\url{https://github.com/harrisonekatz/Learning-Debt-/tree/main}, and the script
is available at
\url{https://github.com/harrisonekatz/Learning-Debt-/blob/main/learning_debt_sim_sensitivity.R}.
The replication package accompanying this submission includes the script, fixed
seeds, replicate-level outputs, summary outputs,
\texttt{table1\_primary\_q75\_v13d.csv}, figures, README, and R session
information. No confidential company data are used in the simulation evidence.
The appendix includes
a proprietary Airbnb production illustration based on monitoring data that cannot
be released. \textbf{Competing interests.} The author is an employee of Airbnb Inc.;
the reproducible simulation does not use Airbnb data.

\clearpage
\appendix
\numberwithin{equation}{section}
\numberwithin{figure}{section}
\numberwithin{table}{section}
\renewcommand{\theequation}{\Alph{section}.\arabic{equation}}
\renewcommand{\thefigure}{\Alph{section}.\arabic{figure}}
\renewcommand{\thetable}{\Alph{section}.\arabic{table}}
\section*{Appendix}

\section{Supplementary v13d figures}
\label{app:simulation-figures}

Figures~\ref{fig:calendar-supp}--\ref{fig:sensitivity-supp} provide additional
views of the deployment-calibrated v13d simulation.

\begin{figure}[htbp]
\centering
\maybeincludegraphics[width=0.92\linewidth]{fig1_relative_to_tuned_calendar_v13d.pdf}
\caption{Utility-calibrated debt policies relative to tuned calendar retraining under the primary 75th-percentile score-unit scaling.}
\label{fig:calendar-supp}
\end{figure}

\begin{figure}[htbp]
\centering
\maybeincludegraphics[width=0.92\linewidth]{fig2_policy_comparison_kappa1_v13d.pdf}
\caption{Policy comparison at $\kappa=1$ under the primary 75th-percentile score-unit scaling.}
\label{fig:kappa1-supp}
\end{figure}

\begin{figure}[htbp]
\centering
\maybeincludegraphics[width=0.92\linewidth]{fig3_no_shift_retrains_v13d.pdf}
\caption{No-shift retraining behavior under the primary 75th-percentile score-unit scaling.}
\label{fig:noshift-supp}
\end{figure}

\begin{figure}[htbp]
\centering
\maybeincludegraphics[width=0.92\linewidth]{fig4_calibration_debt_regret_v13d.pdf}
\caption{Calibration relationship between age-adjusted debt and predictive regret.}
\label{fig:calibration-supp}
\end{figure}

\begin{figure}[htbp]
\centering
\maybeincludegraphics[width=0.92\linewidth]{fig6_score_unit_sensitivity_v13d.pdf}
\caption{Score-unit sensitivity for age-adjusted debt-threshold, debt-utility, and hybrid-utility policies.}
\label{fig:sensitivity-supp}
\end{figure}

\clearpage
\section{Proprietary production illustration}
\label{app:airbnb}

This appendix is included to show why the learning-debt framing is useful in an actual forecasting operation. It is not part of the reproducible simulation evidence. The data come from an internal Airbnb booking lead-time monitoring workflow and cannot be released. For that reason, the case should be read as a production illustration rather than as a reproducible empirical benchmark.

The monitoring path covers 104 weekly observations from January 2023 through December 2024. Each week is summarized as a 12-component booking lead-time composition over future recognition windows. For exposition, the windows are grouped into near, mid, long, and far horizons. The production baseline is a semi-annual calendar retrain. The debt policy uses a scalar additive-log-ratio monitoring statistic, a conjugate NIG shadow posterior, and the same excess-loss decision threshold as the main text. In this illustration, \(\kappa=2\), so the threshold is \(\kappa/(1+\kappa)=2/3\).

The observed platform event is a January 26, 2024 payment-policy shock. The booking composition moves toward near-horizon windows after the shock. The posterior debt signal and the filtered stale probability rise after the event, and the debt policy triggers on March 8, 2024. This is 16 weeks before the next scheduled semi-annual retrain on June 28, 2024. The policy also triggers once before the event, on June 2, 2023, showing that posterior debt can accumulate during a long frozen deployment spell even without an externally visible shock.

Under the same one-step excess-loss accounting used in the paper, the semi-annual calendar policy ends the two-year window at 9.33 excess-loss units, while the debt policy ends at 3.33 units, a ratio of 0.36. The calendar total is \(22\times 1/3 + 3\times 2/3\): 22 stale waiting weeks after the January 2024 shock, plus three unnecessary calendar retrains. The debt-policy total is \(6\times 1/3 + 2\times 2/3\): six stale waiting weeks after the shock, plus two unnecessary retrains. This decomposition is not offered as public benchmark evidence. Its role is to make the operational cost trade-off concrete.

\begin{figure}[htbp]
\centering
\maybeincludegraphics[width=0.92\linewidth]{airbnb_composition.png}
\caption{Proprietary Airbnb production illustration: weekly booking lead-time composition grouped into near, mid, long, and far recognition windows. The red dashed line marks the January 26, 2024 payment-policy shock.}
\label{fig:airbnb-composition-supp}
\end{figure}

\begin{figure}[htbp]
\centering
\maybeincludegraphics[width=0.92\linewidth]{airbnb_kl.png}
\caption{Proprietary Airbnb production illustration: exact KL learning debt between the deployed and shadow NIG monitoring posteriors.}
\label{fig:airbnb-kl-supp}
\end{figure}

\begin{figure}[htbp]
\centering
\maybeincludegraphics[width=0.92\linewidth]{airbnb_rho.png}
\caption{Proprietary Airbnb production illustration: filtered stale probability relative to the \(2/3\) decision threshold. The debt policy triggers on March 8, 2024, before the next scheduled calendar retrain.}
\label{fig:airbnb-rho-supp}
\end{figure}

\begin{figure}[htbp]
\centering
\maybeincludegraphics[width=0.92\linewidth]{airbnb_loss.png}
\caption{Proprietary Airbnb production illustration: cumulative excess loss for the debt policy and the semi-annual calendar policy.}
\label{fig:airbnb-loss-supp}
\end{figure}

\end{document}